# Molecular junctions for thermal transport between graphene nanoribbons: covalent bonding *vs.* interdigitated chains

Alessandro Di Pierro, Guido Saracco, Alberto Fina*

Dipartimento di Scienza Applicata e Tecnologia, Politecnico di Torino, Alessandria Campus,

Viale Teresa Michel 5, 15121 Alessandria, Italy

*Corresponding author: Alberto Fina

e-mail address: alberto.fina@polito.it

## Abstract
Proper design and manufacturing thermal bridges based on molecular junctions at the contact between graphene platelets or other thermally conductive nanoparticles would provide a fascinating way to produce efficient heat transport networks for the exploitation in heat management applications. In this work, using Non Equilibrium Molecular Dynamics, we calculated thermal conductance of alkyl chains used as molecular junctions between two graphene nanoribbons, both as covalently bound and Van der Waals interdigitated chains. Effect of chain length, grafting density, temperature and chain interdigitation were systematically studied. A clear reduction of conductivity was found with increasing chain length and decreasing grafting density, while lower conductivity was observed for Van der Waals interdigitated chains compared to covalently bound ones. The importance of molecular junctions in enhancing thermal conductance at graphene nanoribbons contacts was further evidenced by calculating the conductance equivalence between a single chain and an overlapping of un-functionalized graphene sheets. As an example, one single pentyl covalently bound chain was found to have a conductance equivalent to the overlapping of an area corresponding to about 152 carbon atoms. These results contribute to the understanding of thermal phenomena occurring within networks of thermally conductive nanoparticles, including graphene nanopapers and graphene-based polymer nanocomposites, which are or high interest for the heat management application in electronics and generally in low-temperature heat exchange and recovery.

## Keywords
Thermal Conductance, Molecular Junction, Thermal transport, Non Equilibrium Molecular Dynamics, Graphene.

## Introduction
In the last decade, a great deal of research interest has been devoted to graphene[1-2], owing to its amazing characteristics: electrical properties[3-4], thermal conductivity[5-7], mechanical stiffness and strength[8] which are on the top of the known materials. Since its discover, significant effort has also been devoted to exploring graphene and graphene-related materials[9] for various applications, such as in energy storage[2], nanoelectronics[2], sensors and functional materials[1-2] In several of these fields, thermal conductivity of graphene may play an important role whenever dissipation of heat is crucial.  In particular, in high-performance electronics, heat removal has become a critical issue[7]. Owing to their ultrahigh thermal conductivity, graphene and graphene-related materials have been suggested as candidates in nanomaterials for efficient thermal management.

Significant efforts have been devoted to the preparation of nanocomposite materials embedding graphene and graphene-related materials. However, graphene properties are only partially reflected in nanocomposites[10], where thermal transport is not limited by the thermal conduction of graphene itself but rather by the high thermal resistance at the contact between nanoplatelets[11-13]. In fact, the reach of a percolation threshold, a value over which the particles inside a composite material get physically in contact, is not enough to improve radically the overall thermal conductivity[14]. Indeed, for every single contact, physical limitations in phonon transfer have to be taken into account, leading to a highly inefficient heat transfer across the nanoparticles network that characterize nanocomposites materials[15]. Compatibilization between nanoparticles or between nanoparticles and matrix have been suggested by various authors[16-18] as a possible solution to decrease interfacial resistance. A chemical compatibilization is typically obtained by the functionalization of nanoplatelets with molecules able to build chemical bridges towards the matrix. Different functionalization methods are possible for graphene, either through covalent and non-covalent surface modifications[19]. Covalent functionalization leads to rehybridization of a fraction of the carbon atoms in graphene, thus introducing defects in the $sp^2$ graphene bidimensional lattice[20], creating an alteration of the pristine phonon scatter modes and therefore reducing the conductivity of the nanoplatelets[19]. On the other hand, the non-covalent interactions between the platelets and suitable molecules[19] does not damage the atomic structure, so the thermal proprieties of the graphene layers are preserved. However, the weak forces involved between the molecules can still be a remarkable bottleneck in thermal conductivity[15].

Computational tools, including first principle quantum calculations and Molecular Dynamics (MD), have been crucial to study and evaluate the interfacial properties between graphene platelets as well as between graphene and the matrix in a nanocomposite. In an experimental and computational work, Han *et al*[10] exploited silane functionalization to reduce thermal resistance between substrate and supported graphene nanoflakes. Both computational and experimental results agree that the bridging molecule density tuned the thermal conductivity leading to an heat conduction enhancement. A Reverse Non-Equilibrium MD (RNEMD) approach was proposed from Gao and Müller[21] using bonded fragments of Polyamide 6,6 backbone molecule between graphene sheets. The interfacial and planar thermal boundary conductance (TBC) were evaluated as a function of grafting density and molecule length effect, finding an optimum density value. These results were used in EMA (Effective Medium Approach) to develop a predictive model to extrapolate the macroscopic conductivity of a massive nanocomposite. Liu *et al*[22] used $C_2H_4$ linkers to joint two partially overlapping graphene nanoribbons. Thermal conductance and thermal jump were related to the position and amount of linkers used in the simulations. An inversely proportional relation was found between the thermal jump across the junction and the number of cross-linkers. The related phonon spectra were described as effective mainly on the out of plane vibrational modes due to a mismatch between the in-plane vibrational modes of the linkers and the graphene ones. Chen *et al*[23] simulated the use of external force and chemical functionalization through $CH_2$ linkers across stacked nanotubes to investigate thermal transport, the conductance going through a maximum at the optimal concentration of linkers. A single molecular junction analysis was studied by first-principles full quantum calculations by Li *et al*[24] to evaluate the thermal conductance in short alkanes between two graphene nanoplatelets. The thermal conductance as a function of the molecule stretching state was evaluated, determining that an extended junction almost doubles the thermal conductance compared to the compressed form. In a more recent paper, the same group addressed the comparison between covalently bound molecular junction *vs.* non-covalently bonded π-stacked systems, showing strong suppression of phonon transport, leading to about 95% reduction in thermal conductance for π-stacked junctions, compared to the covalent one[25]. In this work, we investigate the effect of edge grafted short $(CH_2)_n$ alkyl chains as a thermal bridge between two adjacent graphene nanosheets, by the calculation in interfacial conductance via classic Non Equilibrium Molecular Dynamics simulations (NEMD). In particular, the effect of chain length and grafting densities were systematically studied for alkyl chains covalently bound on both graphene nanoribbons. Furthermore, comparisons were done between covalently bonded molecules and interdigitated ones (chains covalently grafted on one graphene sheet and interacting with opposite chains via van der Waals forces), as well as with non-functionalized partially overlapped graphene nanoribbons.

## Theory and Computational Methods / MD modeling

Classical Molecular Dynamics (MD) calculations was used on LAMMPS (Large-scale Atomistic Molecular Massively parallel Simulator) package code witch uses Velocity Verlet integration algorithm to recalculate velocities and positions of the atoms. The 2$^{nd}$ generation of the AIREBO (Adaptive Intermolecular Reactive Empirical Bond Order)[26] force field was employed due to its thermal transport capability as described from a wide literature on carbon and graphene family structures[27-29] and hydrocarbons[30].

The model topology was composed by two ≈100 Å graphene nanoribbons in zigzag configuration joint from alkyl chains chemically grafted between the two graphene sheets, as shown in Figure 1C. The width (Y axis) was set to ≈50 Å, while the total length of simulation box (X axis) varied from 193 Å to 200 Å, depending on the length of the chains, in a stretched configuration, among the sheets. The thickness of the sheet (Z axis) was assumed 3.45 Å from the Van der Waals (VdW) diameter of the carbon atom and the C-C is 1.41 Å as average value, very close to which reported from Diao *et al* [29].

NEMD calculations were performed by the application of Nosé-Hoover thermostats at the two ends of the simulation box, which coincides with the graphene sheets ends. The Hot (310 K) and the cold bath (290 K) of the thermostats regions (last 10 Å) are set as NVT canonical ensemble (constant Number of atoms, Volume and Temperature) while the slabs among the two thermostats ran under NVE (constant Number of atoms, Volume and Energy) condition. All the simulations were carried out for 5 ns using a 0.25 fs timestep. An initial 500 ps of thermal equilibrium at 300 K was set as previously reported by Mortazavi et al[31-32]. A second 500 ps stage of transient non-equilibrium heating, followed with the purpose to reach a constant heat flux. After those initial stages, the constant energy flowing through the thermostats[33] started recording. The thermal flow inside NVE regions is derived from the slope of energy versus time plots[31]. Additional investigations involved different equilibrium and run temperatures: 200 K; 400 K; and 500 K while the ΔT at the thermostats was kept constant to ±10 K as previously described. In this range the temperature quantum correction suggested at lower temperatures by classic Molecular Dynamic simulation is considered neglibile[34].

To evaluate the temperature variation inside the sample along the thermal flow, the simulation box length was virtually split transversally into 22 thermal layers. The temperature of each thermal layer was then computed using the relation in Equation 1,

$$T_i(slab) = \frac{2}{3N_i k_B} \sum_j \frac{p_j^2}{2m_j} \qquad (1)$$

In Equation 1, $T_i$(slab) is the temperature of $i^{th}$ slab, $N_i$ is the number of atoms in $I^{th}$ slab, $k_B$ is the Boltzmann's constant, $m_j$ and $p_j$ are atomic mass and momentum of atom *j*, respectively. All the slabs temperatures were then time-averaged to the simulation runtime excluding the non-linear regions at the nanoribbon boundary, *i.e.* the ones close to the thermostats and the ones coupled to the junction. The thermal boundary conductance (TBC) of the model, *G,* expressed in pW/K, was evaluated using the Equation 2.

$$G = \frac{q_x}{\Delta T} \qquad (2)$$

Here $q_x$ is the thermal flow derived from the energy versus time plot slope and $\Delta T$ is the thermal difference across the jump, as projection of the two linear fit of the temperature-length plot in the junction middle point.

The single chain thermal conductance, G$_{chain}$, expressed in pW/K was calculated through the equation 3.

$$G_{chain} = \frac{q_x}{n \cdot \Delta T} \qquad (3)$$

where n is the number of molecules bridging the two adjacent graphene nanoribbons.

## Molecular Dynamics modeling and results

To study the effect on thermal properties of organic chains bridging between two adjacent nanoribbons, MD calculation were carried out as function of alkyl chain length, grafting density and type of bonding (covalent or Van der Waals interaction) between the organic chains and the graphene nanoribbons.

As a first parameter evaluated, the length of the alkyl chains connecting the two graphene sheets via covalent bonding on both nanoribbons was addressed as illustrated in Figure 1A-A'. Five different backbones chain length were tested (in brackets the short name and the molecule length in elongated conformation): methyl (C1; 2.46 Å), propyl (C3; 4.97 Å), pentyl (C5; 7.49 Å), eptyl (C7; 10.00 Å) and tridecyl (C13, 18.53 Å), with a constant grafting density, defined as the number of chains per unit length of graphene armchair edges. The second parameter addressed is the grafting density, while keeping constant the length of the covalently bound chains. The maximum grafting density in this work was defined as one chain per aromatic ring along the armchair graphene edge (4.26 Å spacing distance), referred to as 111. Lower grafting densities were defined as one chain every two aromatic rings, referred to as 101, and one chain every 3 aromatic ring, referred to as 100, as depicted in Figure 1A-A''. In the third part of this work, alkyl chains covalently bound on a single side were addressed, producing an interdigitated structure (Figure 1B) of alkyl chains grafted alternatively on the two nanoribbons, aiming at the study of thermal conductance via Van der Waals forces between the chains, in comparison with covalent bonding of chains on both nanoribbons.

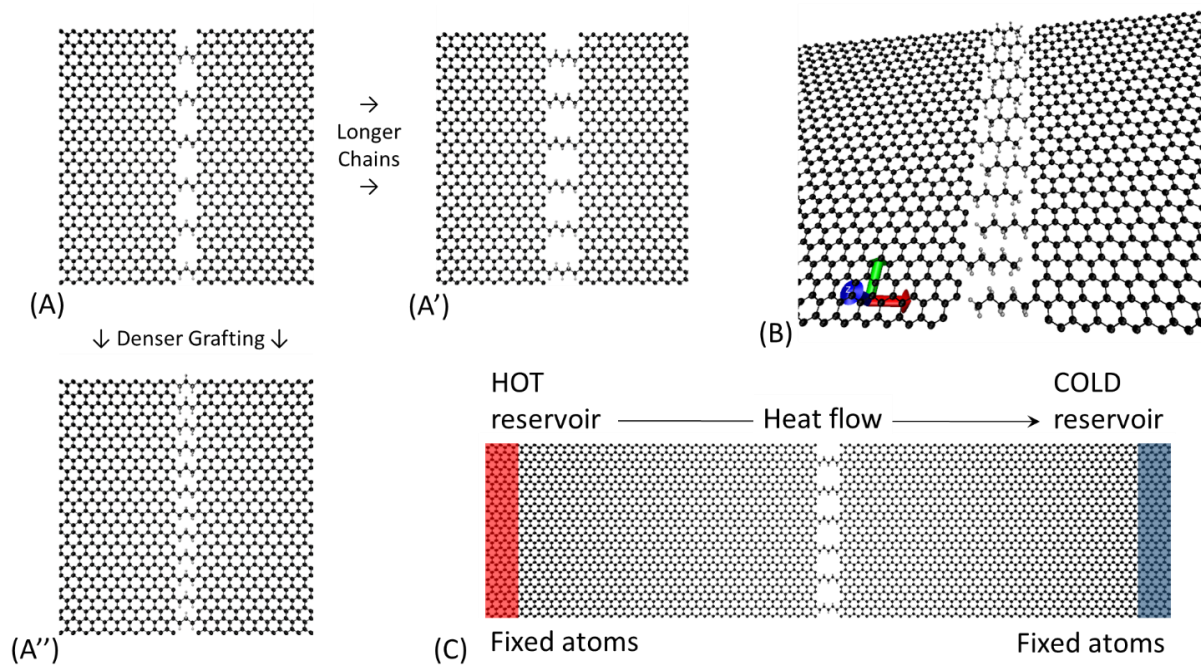

**Figure 1.** Molecular junctions graphical description. A) The 101 C3 model top view shows for the C3 chain length an alternate configuration between chains and empty sites while A') represents the same configuration for the longer C5 chain length and A'') shows denser grafting scheme 111 is for the C3 chain length. B) The interdigitated model junction detail with pending pentyl chains. C) The top view Molecular Dynamics model for the evaluation of thermal conductance with highlighted reservoirs. VMD software is used for all graphical representations[35].

The Energy added to the hot reservoir and removed from the cold reservoir versus time were found linear and symmetric, with slopes dependent on chain length and grafting density. These observations evidence that the total energy inside the ensemble is kept constant and a constant heat flux passes through the model. Plots obtained for covalently bound chains C1 and C13 with the maximum grafting density are reported in Figure 2A, as an example. Quantitatively, the slopes varied from 0.19 eV/ps for the longer C13 chain to 0.40 eV/ps for the shortest C1 chain (see Figure S1 for other chain lengths and grafting densities). However, dependency of the slope with the chain length is not linear, as the slope is almost constant for C5 chains or longer. Heat flux, $q_x$, defined as the energy transferred from the atoms inside the hot reservoir to those inside the cold reservoir at each timestep, was calculated from these curves.

The temperature profiles along the graphene nanoribbons and across the interface between those are reported in Figure 2B for C1 and C13, 111 grafted chains, taken as an example. Despite a non-perfect convergence to the thermostats value due to the exclusion of the boundary regions which classically shows non-linearity close to the thermostats[32], linear fitting of the temperatures along the graphene sheets allow to extrapolate the temperature jump across the interface, which is a function of the length of the bridging chain (see Figure S2 for all chain lengths and grafting densities). As expected, shorter chain bonding exhibits a lower thermal jump between the graphene sheets compared to the longer ones, reflecting a better thermal exchange through the junction, *i.e.* a higher TBC, accordingly with Equation 2. Similar temperature plots are obtained for all of the chain length explored and the three different grafting schemes. Temperature jump dependency on grafting density is in qualitative agreement with previous report form Liu *et al*[22] in partially overlapped nanoribbons.

Thermal boundary conductance was calculated from equation 2 and the related plot of the values is reported in Figure 2C. The TBC rapidly decreased with increasing length up to C5 chain, whereas TBC for longer chains appear to level off around 150 pW/K. In the diffusive regime, a decrease in the thermal conductance is obviously expected with increasing chain length. In the case of an ideal material observing Fourier's law, the conductance should decrease linearly with increasing the chain length. However, it was previously reported that the decrease in conductance of alkyl chains is not linear with the chain length[36]. Examples of asymptotic decrease of conductance with increasing alkyl chain length were reported by several authors, both from experimental measurements[36] and computational work[37-38]. Different explanations have been proposed, including quasi-ballistic transport, effect of direct mechanical coupling or the variation of phonon modes as a function of chain length[36]. Figure 2C also reports the effect of the different grafting schemes: as expected, the reduction in the number of bridging chains per unit length leads to a rapid decay of the interfacial thermal conductance. Table 1 reports $G_{chain}$ values calculated from Equation 3 for all the cross-linkers schemes and chain lengths. $G_{chain}$ values are found independent on the grafting scheme and therefore reflect the contribution of single chains to the interface between two nanoribbons, confirming the additive effect of the single linker reported in previous works[10, 22]. Quantitatively, conductance values are aligned with results previously reported in the literature. Indeed, Li *et al.*[24] reported a $G_{chain}$ of about 160 pW/K at 300K for edge-grafted C11 alkane chain bridging two graphene nanoribbons, while $G_{chain}$ of about 330 pW/K was calculated for -$CH_2$-$CH_2$- linkers intercalated between graphene planes[22]. Conductance values of the same order of magnitude (about 250 pW/K) were also reported for methylene linkers between carbon nanotubes[23]. Figure 2D reports the TBC as a function of the equilibrium temperature in the range between 200 and 500 K. Limited variations (± 10% on the average values) were obtained and no clear trend was observed for thermal conductance values with increasing temperature. We therefore assume thermal conductance approx. constant in the explored temperature range.

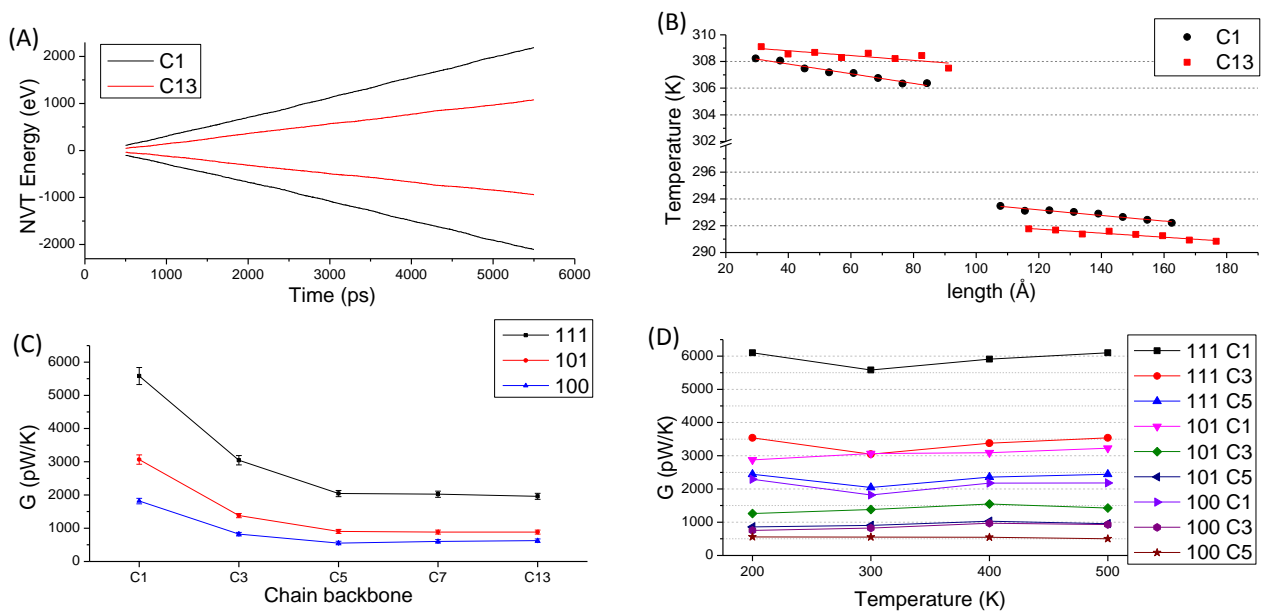

**Figure 2.** (A) Energy added to the hot reservoir (positive slope) and removed from the cold reservoir (negative slope) for C1 and C13 chains, as an example. (B) temperature trend along the nanoribbons for the shortest C1 and the longest C13 chain lengths. (C) Thermal boundary conductance as a function of the backbone chain length, each series represent a different grafting scheme. The lines guide the eye on the chain length effect. Alternatively, a vertical reading direction highlights the thermal conductance as a function of grafting density. The TBC uncertainty is estimated from variations obtained between simulations made using three different seeds per run in velocities assignments, on selected chain length grafting configurations (See Table S1 in supporting information for details). (D). Dependency of thermal conductance for C1, C3 and C5 chain lengths in all grafting schemes at 200 K, 300 K, 400 K and 500 K.

**Table 1.** Average single chain conductance ($G_{chain}$) for covalently bound chains.

| Chain Backbone | 111 $G_{chain}$ (pW/K) | 101 $G_{chain}$ (pW/K) | 100 $G_{chain}$ (pW/K) | Average $G_{chain}$ (pW/K) |
|---|---|---|---|---|
| **C1** | 455 | 504 | 442 | 467 ± 31 |
| **C3** | 252 | 230 | 215 | 232 ± 19 |
| **C5** | 169 | 149 | 137 | 152 ± 16 |
| **C7** | 169 | 147 | 151 | 156 ± 11 |
| **C13** | 163 | 147 | 156 | 156 ± 8 |

Beside the effect of chains covalently bound on two adjacent nanoribbons, in this work the effect of chains covalently bound on one graphene sheet and interacting via VdW bonding with chains attached on another graphene sheet (Figure 1B) was also addressed. In fact, from the application point of view, this scenario is much more relevant to the interaction between functionalized graphene nanosheets, provided functionalization is able to induce a self-assembly of nanoplatelets. In the study of VdW interactions, the chain length was fixed as C5, while the equilibrium distances between the nanoflakes was stepwise increased at 9.8, 10.6, 11.5, 12.4 and 13.0 angstrom, directly reflecting a decrease in the interdigitation depth between chains, as depicted in Figure 3.

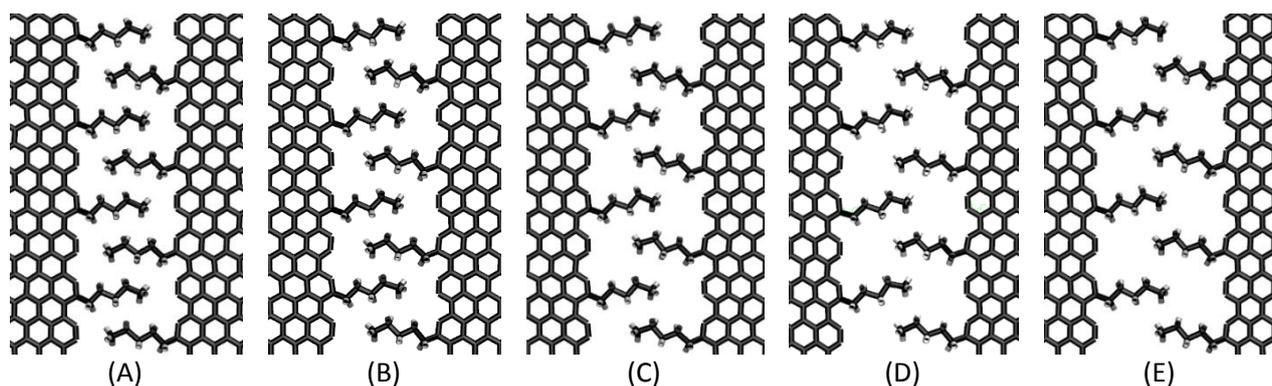

(A) (B) (C) (D) (E)

**Figure 3**. Interface detail of interdigitated models with various distance between nanoribbons: (A) 9.8 Å, (B) 10.6 Å, (C) 11.5 Å, (D) 12.4 Å and (E) 13.0 Å.

Figure 4 depicts the single chain conductance of the interdigitated models as a function of the platelets distance. The single chain thermal conductance exhibits a linear decrease, from 131 to 24 pW/K when the nanoribbons distance increases from 9.8 to 13.0 Å, which clearly depends on the extent of interaction surface between interdigitated molecules. It is worth noting that conductivity values are lower than the primary bonded C5 chain (152 pW/K) evaluated previously, because of the weaker interaction via VdW forces compared to covalent bonding.

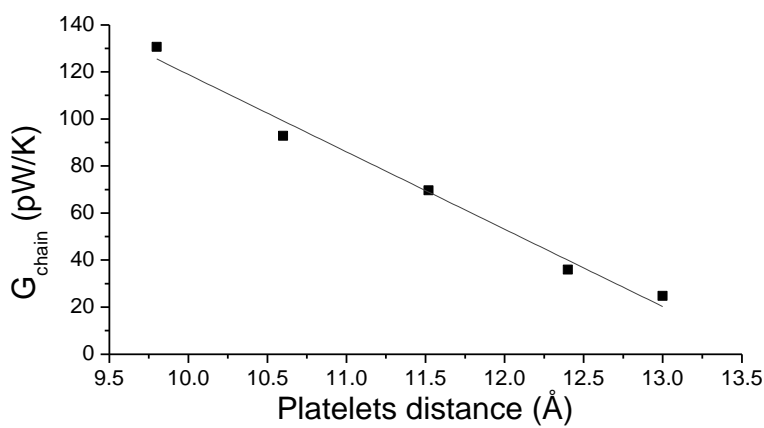

**Figure 4.** Single chain thermal conductance of interdigitate models as a function of platelets distance.

A set of partially overlapped graphene nanoribbons with the same dimensions as the set used before (about 100 Å in length x 50 Å in width) was simulated to compare the value of their surface specific interfacial conductance as a function of the overlapped surface (Figure 5) The platelets distance (3.40 Å) and offset between layers were set corresponding to the actual graphite crystal lattice stacking. Overlapped area was varied from one-half, to one-third and one-quarter of the nanoribbons surface, corresponding to 4395 Å$^2$, 2973 Å$^2$ and 2198 Å$^2$, respectively. Thermal conductance for these overlapped areas therefore calculated as 1681, 1024 and 907 pW/K, respectively, yielding to an average of 0.38±0.04 pW/Å$^2$K. Based on this value, an equivalence in conductance was obtained between a single C5 chain (152 pW/K) and an area of 400±42Å$^2$, corresponding to 152±16 atoms per layer, assuming the previously reported[39] atomic density value of 0.38 atoms/Å$^2$. Such a comparison provides an estimation of correspondence between TBC through simple graphene overlapping and trough covalent crosslinking.

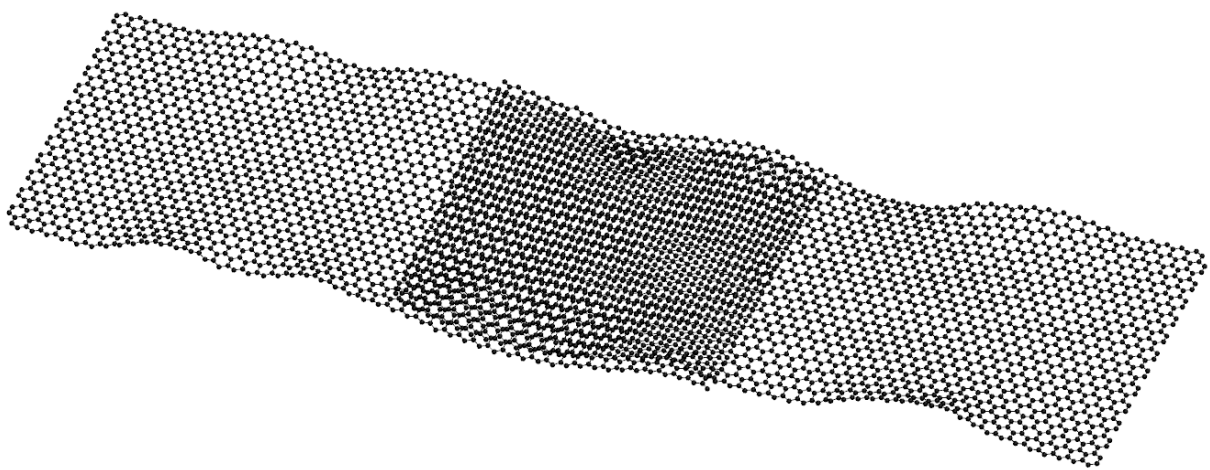

**Figure 5.** A one third partial overlapped nanoribbons, corresponding to a contact area of 2973 Å$^2$ in VMD[35] orthographic view.

## Conclusions

The thermal boundary conductance between graphene nanosheets bonded by covalent molecular junctions or via VdW interdigitated chains in head-head configuration was evaluated using classical non-equilibrium Molecular Dynamics simulations. The role of differently bonded aliphatic chains of different length, grafting density and bond type was evaluated. The simulation results highlighted a strict dependence of the TBC on the bridging molecule length. Additionally, the individual thermal conductance for the alkyl chains was found also in the range of 10$^2$ pW/K. The highest conductance was obtained for shorter chains (up to about 470 pW/K for C1 junction) with a dramatic decay with the chain length increase, until a conductance plateau is reached (at about 150 pW/K). Moreover, the grafting density for this model was found linearly correlated to the interfacial thermal conductance. The analysis of Van der Waals bonded interdigitated molecules revealed a lower thermal conductance compared to covalent bonded ones, linearly decreasing as a function of the platelets distance, *i.e.* the reduction in molecules interdigitation. Equivalence in terms of conductance was also calculated for partially overlapping non-functionalized graphene, showing equivalence between a single C5 chain with an overlapped area of 400±42 Å$^2$, which corresponds to 152±16 carbon atoms per layer. Covalent chain bridging is therefore a rather efficient way, yet experimentally challenging, to produce molecular junction aimed at the efficient heat exchange into a percolating network of graphene nanosheets, to be exploited in graphene nanopapers as well as in graphene-based polymer nanocomposites.

## Authors' contributions

A. Fina conceived the experiments, interpreted results and coordinated the project, A. Di Pierro carried out MD simulations and post processing, G. Saracco contributed to the discussion of the results. Manuscript was written by A. Di Pierro and A. Fina. All the authors have approved the final article.

## Acknowledgements

This work has received funding from the European Research Council (ERC) under the European Union's Horizon 2020 research and innovation programme grant agreement 639495 — INTHERM — ERC-2014-STG.

The authors gratefully acknowledge Dr. Bohayra Mortazavi at Bauhaus University Weimar (D) for continuous advices and support during the preparation of this work. Dr. Diego Martinez Gutierrez at Politecnico di Torino is also acknowledged for the useful discussions.

# SUPPORTING INFORMATION:

# Molecular junctions for thermal transport between graphene nanoribbons: covalent bonding vs. interdigitated chains

Alessandro Di Pierro, Guido Saracco, Alberto Fina*

Dipartimento di Scienza Applicata e Tecnologia, Politecnico di Torino, Alessandria Campus, Viale Teresa Michel 5, 15121 Alessandria, Italy.

*Corresponding author: Alberto Fina; e-mail address: alberto.fina@polito.it

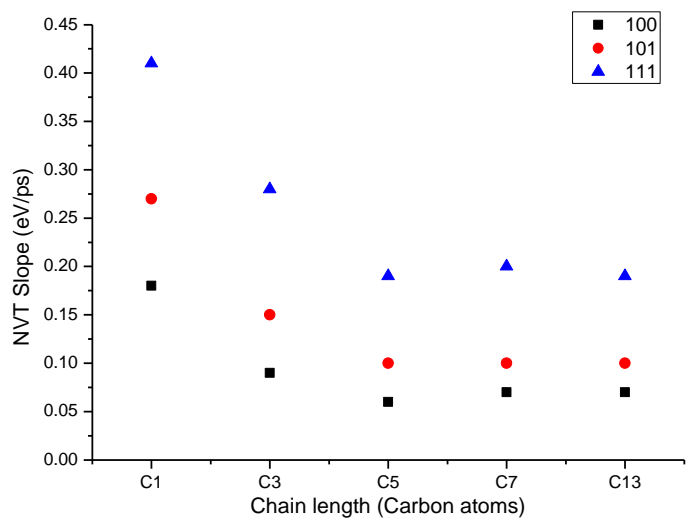

*Figure S1. Energy versus time plot slopes for 300K simulation set. All the three grafting schemes are represented.*

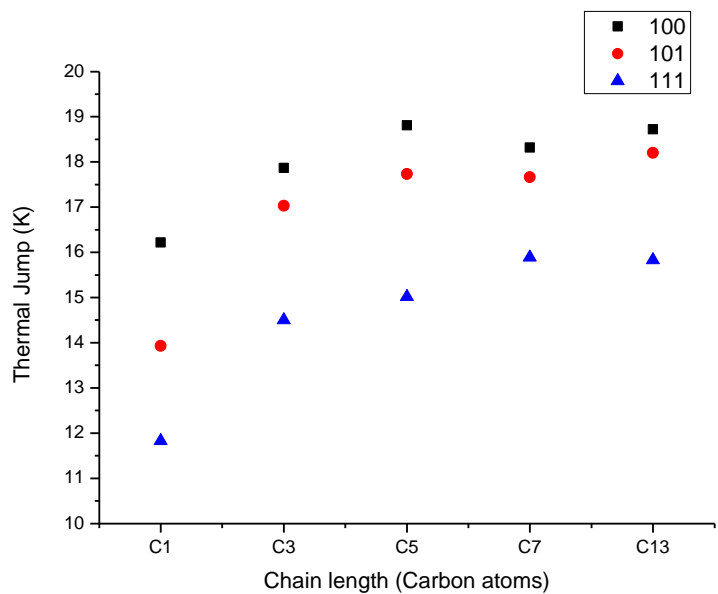

*Figure S2. Thermal jumps as a function of chain length for 300K simulation set. All the three grafting schemes are represented.*

*Table S1. Thermal Boundary Conductances and single chains thermal conductances, energy versus time slopes and thermal jumps for three different velocity seeds at 300 K: average values and deviations.*

| Grafting density C3 chain | TBC [pW/K] | | | | |
|---|---|---|---|---|---|
| | Seed 0 | Seed 1 | Seed 2 | Average | Deviation |
| 111 | 3045 | 3150 | 3200 | 3132 | 77 |
| 101 | 1379 | 1366 | 1431 | 1392 | 33 |
| 100 | 820 | 757 | 885 | 821 | 64 |

| Grafting density C3 chain | $G_{chain}$ [pW/K] | | | | |
|---|---|---|---|---|---|
| | Seed 0 | Seed 1 | Seed 2 | Average | Deviation |
| 111 | 254 | 263 | 267 | 261 | 7 |
| 101 | 230 | 228 | 238 | 232 | 6 |
| 100 | 205 | 215 | 221 | 214 | 9 |

| Grafting density C3 chain | NVT slope [eV/ps] | | | | |
|---|---|---|---|---|---|
| | Seed 0 | Seed 1 | Seed 2 | Average | Deviation |
| 111 | 0.280 | 0.270 | 0.270 | 0.273 | 0.005 |
| 101 | 0.150 | 0.140 | 0.150 | 0.147 | 0.005 |
| 100 | 0.090 | 0.100 | 0.100 | 0.097 | 0.005 |

| Grafting density C3 chain | Thermal Jump [K] | | | | |
|---|---|---|---|---|---|
| | Seed 0 | Seed 1 | Seed 2 | Average | Deviation |
| 111 | 14.50 | 13.59 | 13.64 | 13.91 | 0.46 |
| 101 | 17.03 | 16.98 | 16.79 | 16.93 | 0.12 |
| 100 | 17.87 | 18.05 | 18.09 | 18.00 | 0.11 |